\documentclass{osa-article}

\journal{osajournal}

\articletype{Research Article}

\usepackage{lineno}
\usepackage{color,soul}
\begin{document}

\title{4-Dimensional Symmetry Breaking of Light in Kerr Ring Resonators}

\author{Lewis Hill,\authormark{1,2,*} Pascal Del'Haye,\authormark{2,3} and Gian-Luca Oppo\authormark{1}}

\address{\authormark{1}SUPA \& CNQO, Department of Physics, University of Strathclyde, 107 Rottenrow, Glasgow, G4 0NG, UK\\
\authormark{2}Max Planck Institute for the Science of Light, Staudtstr. 2, 91058 Erlangen, Germany\\
\authormark{3}Department of Physics, Friedrich Alexander University Erlangen-Nuremberg, 91058 Erlangen, Germany}

\email{\authormark{*}lewis.hill@strath.ac.uk} 



\begin{abstract*}
Symmetry breaking of light states is of interest for the understanding of nonlinear optics, photonic circuits, telecom applications and optical pulse generation. Here we demonstrate 4-fold symmetry breaking of the resonances of ring resonators with Kerr nonlinearity. This symmetry breaking naturally occurs in a resonator with bidirectionally propagating light with orthogonal polarization components. The four circulating field components are shown to exhibit multiple and, nested and isolated, spontaneous symmetry breaking bifurcations, and are also capable of complex oscillatory dynamics - such as four-field self-switching, and unusual temporal cavity soliton-like dynamics on time scales of multiple round-trip times, with extended delays between subsequent generations.
\end{abstract*}

\section{Introduction}
Spontaneous symmetry breaking (SSB) phenomena are of fundamental importance to many areas of science with some notable examples being the Higgs mechanism in particle physics \cite{bernstein1974spontaneous}, the superconductivity of metals in condensed-matter physics, early universe models \cite{kazanas1980dynamics}, plasmonics \cite{barbillon2020applications}, even the evolution of swimming and flying organisms in fluid dynamics and biology \cite{bagheri2012spontaneous}. In the field of non-linear optics, there has been a recent explosion of work studying SSB in Kerr ring resonators.

\begin{figure}
    \centering
    \includegraphics[width=8.5cm]{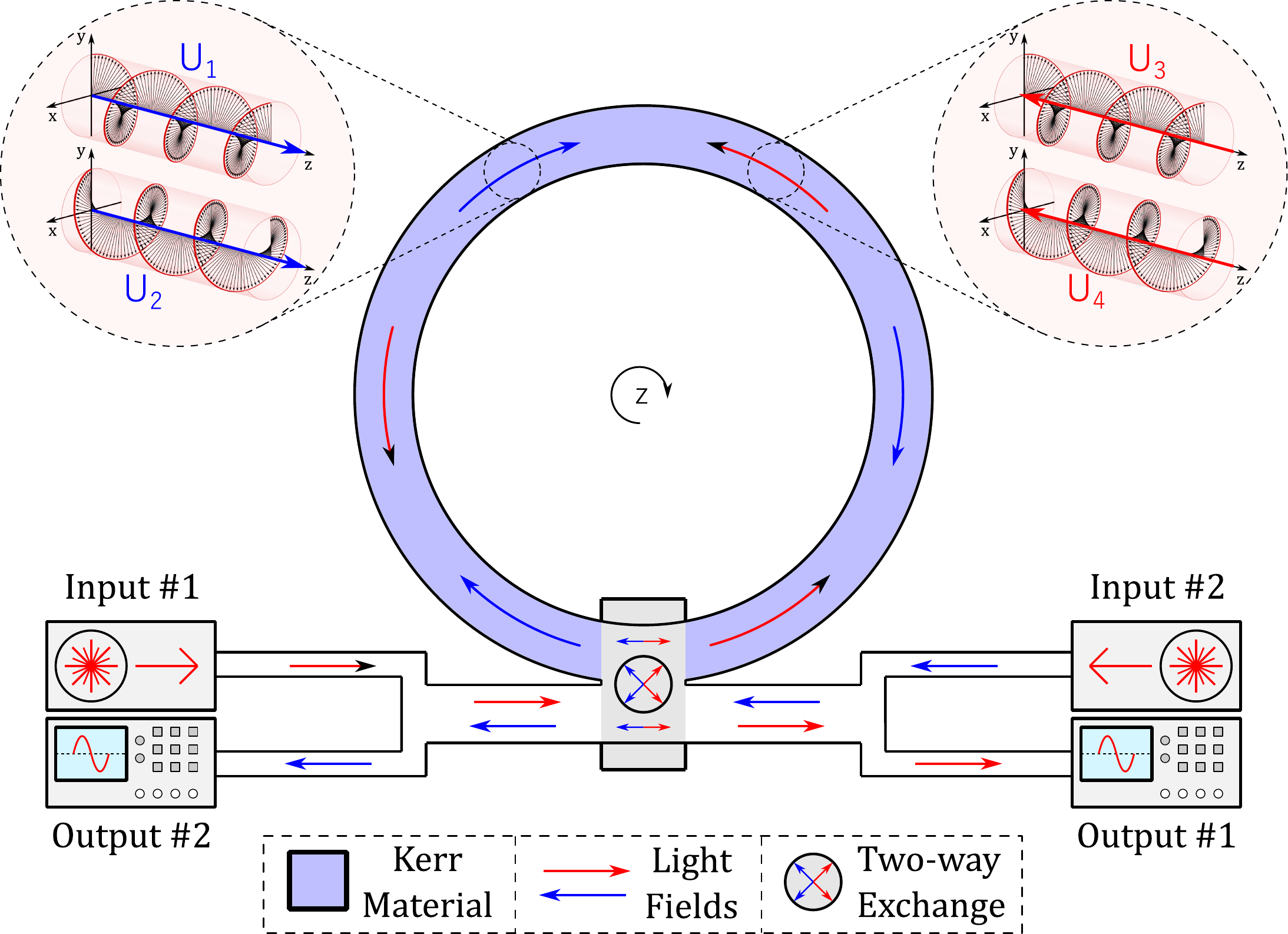}
    \caption{Studied setup. Two identical, linearly polarised, laser inputs enter a Kerr ring resonator (shaded blue) via a two-way coupling mechanism from opposing directions. This leads to two counter-propagating, linearly polarised, fields circulating the resonator. By splitting the linearly polarised light fields into left- and right-circularly polarised components we obtain a total of four circulating fields.}
    \label{fig:Schematic}
\end{figure}

A Kerr ring resonator is a closed loop optical path, made of a Kerr material where the refractive index depends on the intensity of a strongly interacting coherent (laser) field. Laser light enters and leaves the closed path of the resonator through optical couplers such as beam splitters or evanescent coupling, in setups similar to that displayed in Fig. \ref{fig:Schematic}. The cavity fields slowly evolve as they circulate the resonator due to a combination of the input pump powers, laser detunings, interactions with the Kerr material, and losses. Depending on the transmission of the two-way coupler, the fields can circulate within the resonator for a very large number of round trips, allowing for long evolution times. Upon leaving the resonator, via the coupling mechanism, the circulating fields then progress on towards different outputs where they can be further processed or measured. Note however that Fig. \ref{fig:Schematic} shows a more complex scenario than a typical single input Kerr resonator, in that two counter-propagating laser inputs are used rather than a single one. These two inputs are subsequently split into different orthogonal polarisation components. This configuration is the device of reference for the work presented here.

The schematic of the ring resonator in Fig. \ref{fig:Schematic} highlights a combination of two fundamental principles to achieve symmetry breaking. The first symmetry breaking principle is due to the counter-propagation of two input beams in opposite directions, which are otherwise identical. The evolutions of counter-propagating fields circulating ring resonators, including their SSB, have been studied extensively theoretically \cite{kaplan1981enhancement, kaplan1982directionally,wright1985theory, woodley2018universal, bitha2021complex, hill2020effects, woodley2021self}, and, more recently, experimentally \cite{del2017symmetry,del2021optical, silver2017nonlinear}. The second symmetry breaking principle that is important for our discussions is that of a single, linearly polarised, laser input splitting into two orthogonally polarised components. Theoretically this SSB was investigated in \cite{geddes1994polarisation} when considering transverse diffraction via two coupled Lugiato-Lefever equations (LLE) \cite{lugiato1987spatial}, and in \cite{haelterman1994polarization} when considering longitudinal dispersion via two driven and damped nonlinear Schr\"odinger equations. This two-polarization setup, and its SSB, have also seen a recent flurry of studies both theoretical and experimental \cite{woodley2018universal, hill2020effects, garbin2020asymmetric, moroney2022kerr, xu2021spontaneous, xu2022breathing}.

The two separate setups, counter-propagation and two-polarization, are already proving promising in their scope of applications. The polarisation system can be used as a polarization controller \cite{moroney2022kerr} and for producing SSB temporal cavity solitons \cite{xu2021spontaneous} and breathers \cite{xu2022breathing} with the potential to provide novel methods for two-components frequency combs. The counter-propagating system can be used to enhance the Sagnac effect \cite{kaplan1981enhancement} and to realise isolators and circulators \cite{del2018microresonator} for, for example, all optical computing. The motivation behind our work is then to greatly expand on the number of degrees of freedom by combining the two separate systems into a single one. This leads to a system not only capable of replicating many of the above mentioned applications, and often enhancing them, but further allowing for new regimes and applications to be realised such as polarization dependant isolators and circulators.

We consider here the case of spatially homogeneous fields along the resonators and focus on their steady states and dynamics over several round trip times. This manuscript is structured as follows. In section 2, we outline the derivation of our system to describe the field components presented in Fig. \ref{fig:Schematic} before, in section 3, discussing the results of our simulations of the system through use of numerical integration techniques, focusing first on the stationary states of the system and later its possible oscillatory dynamics, section 4 - 5.

\section{Setup and Model}

We study the system that results from two counter-propagating, linearly polarised, laser inputs, which enter the Kerr ring resonator via optical couplers (e.g. a waveguide close to the resonator). The circulating light fields within the resonator can then be decomposed into orthogonal polarization components, as shown in Fig. \ref{fig:Schematic}.

To model this situation we consider first the propagation of a single vectorial field ${\textbf{\textit{E}}}$ in a Kerr ring cavity as discussed in \cite{geddes1994polarisation}. When neglecting dispersion or diffraction, one obtains:

\begin{equation}\label{GedLLENoDisp}
        \frac{\partial\textbf{\textit{E}}}{\partial t} = \textit{\textbf{E}}_{0}-\left(1+i\eta\theta\right)\textbf{\textit{E}}+i\eta\left[A\left(\textbf{\textit{E}}\cdot \textbf{\textit{E}}^\textbf{\textit{*}}\right)\textbf{\textit{E}}+B\left(\textbf{\textit{E}}\cdot \textbf{\textit{E}}\right)\textbf{\textit{E}}^\textbf{\textit{*}}\right],
\end{equation}

\noindent where $\textbf{\textit{E}}=\mathcal{E}_x \hat{\textbf{\textit{x}}} + \mathcal{E}_y \hat{\textbf{\textit{y}}}$ is the normalised vector electric field envelope (comprised of components along the $x$ and $y$ axis respectively, with the cavity axis along $z$), $\textit{\textbf{E}}_{0}$ is the input field, $\eta=\pm1$ in the case of a self-focusing and self-defocusing medium respectively, $\theta$ is the cavity detuning (difference between the input laser frequency and the closest cavity resonant frequency) and the constants $A,B$ represent the self- and cross-phase modulation strengths respectively, which are here given by:

\begin{equation}
    A=\left(\chi_{1122}^{(3)}+\chi_{1212}^{(3)}\right),\;\;\;B=\chi_{1221}^{(3)},
\end{equation}

\noindent where $\chi^{(3)}$ is the third order susceptibility tensor \cite{boyd2020nonlinear}.

The self- and cross-phase modulation constants describe the strengths with which the two polarisation components affect themselves and each other, respectively, as they circulate the ring resonator.

We now seek to generalise this model further by taking into account an additional input which causes a second, counter-propagating, field to circulate within the resonator. We do this by considering the propagation of light in the medium in a manner similar to that outlined in \cite{pitois2001nonlinear}, that is, we set clockwise, subindex $cw$, and counter-clockwise, subindex $ccw$, polarization components via:

\begin{equation}\label{CountPropExp}
    \mathcal{E}_{x,y}=\mathcal{E}_{cwx,cwy}e^{ikz}+\mathcal{E}_{ccwx,ccwy}e^{-ikz}.
\end{equation}

\noindent Here $z$ is the propagation direction along the ring, $k$ the light wavevector, and $(x,y)$ the plane perpendicular to propagation as mentioned above. By expanding Eq. \eqref{GedLLENoDisp} with Eq. \eqref{CountPropExp}, neglecting the fast varying terms, and separating all $\mathcal{E}_{cwx}$, $\mathcal{E}_{cwy}$, $\mathcal{E}_{ccwx}$, $\mathcal{E}_{ccwy}$ terms as far as possible, one eventually arrives at a relatively long system of four coupled equations. We then simplify these equations by moving to a circular polarisation basis defined in Table \ref{table:4Fields}.

\begin{table}
\begin{center}
\begin{tabular}{ p{22mm}|p{30mm}|p{30mm}}
    \; &\begin{center}
        Clockwise propagating
    \end{center}&\begin{center}
        Counterclockwise propagating
    \end{center}\\
    
    \hline
    
    \begin{center}
        Left Circ. Polarised
    \end{center}&\begin{center}
        $U_1=\frac{\mathcal{E}_{cwx}+i\mathcal{E}_{cwy}}{\sqrt{2}}$
    \end{center}& \begin{center}
        $U_3=\frac{\mathcal{E}_{ccwx}+i\mathcal{E}_{ccwy}}{\sqrt{2}}$
    \end{center}\\
    
    \hline
    
    \begin{center}
        Right Circ. Polarised
    \end{center}&\begin{center}
        $U_2=\frac{\mathcal{E}_{cwx}-i\mathcal{E}_{cwy}}{\sqrt{2}}$
    \end{center}&\begin{center}
        $U_4=\frac{\mathcal{E}_{ccwx}-i\mathcal{E}_{ccwy}}{\sqrt{2}}$
    \end{center}
\end{tabular}
\caption{Transformations to circular polarisation basis}
\label{table:4Fields}
\end{center}
\end{table}

In this basis our model takes a more succinct form:

\begin{equation}\label{CircPolBasisModel}
    \begin{split}
        \frac{\partial U_1}{\partial t} = E_{in}&-\left(1+i\eta\theta\right)U_1 +i\eta\left[\left(A|U_1|^2+C|U_2|^2+2A|U_3|^2+C|U_4|^2\right)U_1 +CU_3U_4^*U_2\right],\\
        \frac{\partial U_2}{\partial t} = E_{in}&-\left(1+i\eta\theta\right)U_2 +i\eta\left[\left(A|U_2|^2+C|U_3|^2+2A|U_4|^2+C|U_1|^2\right)U_2 +CU_4U_3^*U_1\right],\\
        \frac{\partial U_3}{\partial t} = E_{in}&-\left(1+i\eta\theta\right)U_3 +i\eta\left[\left(A|U_3|^2+C|U_4|^2+2A|U_1|^2+C|U_2|^2\right)U_3 +CU_1U_2^*U_4\right],\\
        \frac{\partial U_4}{\partial t} = E_{in}&-\left(1+i\eta\theta\right)U_4 +i\eta\left[\left(A|U_4|^2+C|U_1|^2+2A|U_2|^2+C|U_3|^2\right)U_4 +CU_2U_1^*U_3\right],
    \end{split}
\end{equation}

\noindent where $E_{in}$ is the input amplitude $\textit{\textbf{E}}_{0}$ in this basis and $C=A+2B$. Similar to Eq. \eqref{GedLLENoDisp}, the first terms within the square brackets of Eqs. \eqref{CircPolBasisModel} are caused by self-phase modulation, whereas the second, third, and fourth terms are caused by cross-phase modulation. The last term is responsible for an energy exchange between the two circular components of each beam \cite{pitois2001nonlinear}, an exchange which is not present in the separate models. This final component also prevents us from finding the homogeneous stationary states (HSS) of the system in the usual manner \cite{hill2020effects}, although we can still integrate Eqs. \eqref{CircPolBasisModel} to search for stable or dynamic solutions. For equal pumping and detunings, system (4) is invariant under the transformations that exchange the indexes 1 with 2 and 3 with 4 (polarization component exchange), that exchange the indexes 1 with 3 and 2 with 4 (line counterpropagation component exchange) and that exchange the indexes 1 with 4 and 2 with 3 (cross counterpropagation component exchange). These are the symmetries that will be broken by the bifurcations described in the next sections.

\section{\label{sec:ComSim}Simulating the Evolving Field Intensities}

For the rest of the article and in the numerical simulations of Eqs. \eqref{CircPolBasisModel}, we use values of $A=2/3$ and $C=4/3$ since these correspond to those associated with silica glass fibers, where $\chi_{1122}^{(3)}\approx\chi_{1221}^{(3)}\approx\chi_{1212}^{(3)}\approx\chi_{1111}^{(3)}/3$.

\begin{figure}
    \centering
    \includegraphics[width=8.5cm]{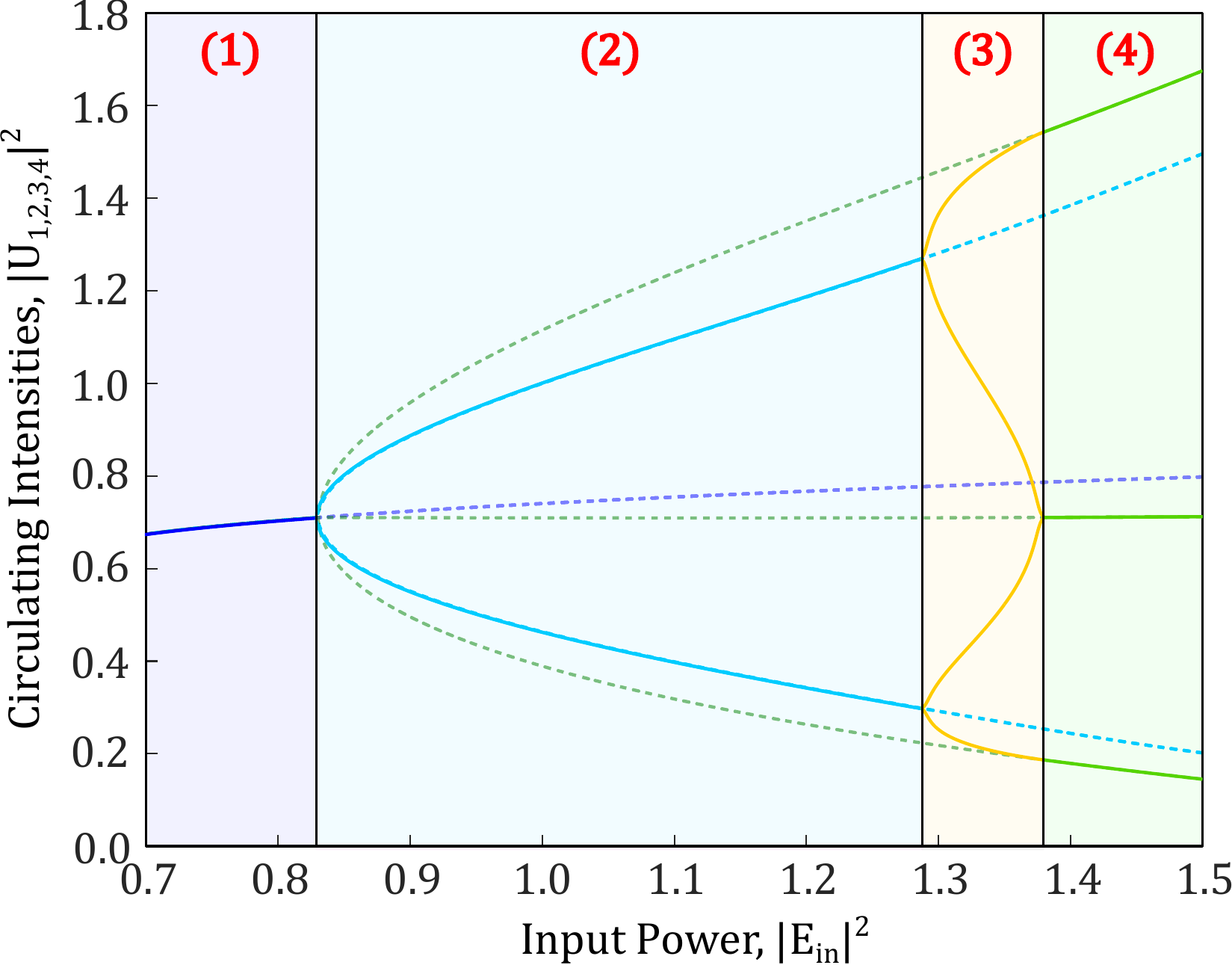}
    \caption{An input power scan of Eq. \eqref{CircPolBasisModel} for $\theta=3.85$. The scan runs from right to left with the previous end iteration used as the initial condition for the next input value. The solid lines shows the results with Eq. \eqref{CircPolBasisModel} having no constraints, i.e those which are attractive and stable?. The dashed green line shows Eq. \eqref{CircPolBasisModel} with forced $U_1=U_4$, the dashed cyan line has forced $U_1=U_2$ and $U_3=U_4$, and the dashed blue line has forced $U_1=U_2=U_3=U_4$. Note that the solid line has four distinct regions with shaded backgrounds. Region (1) has full symmetry where all fields have equal intensity; region (2) begins with a symmetry breaking pitchfork bifurcation and results in two asymmetric pairs of symmetric fields; region (3) begins with a second symmetry breaking pitchfork bifurcation, leading to four entirely asymmetric fields, and finally region (4) begins with a partial-symmetry restoring pitchfork bifurcation, leading to a single pair of symmetric fields, with the remaining two fields being both asymmetric to each other and to the symmetric pair.}
    \label{fig:NestedSym}
\end{figure}

Figure \ref{fig:NestedSym} shows the four circulating field intensities during an input intensity scan for a cavity detuning value of $\theta = 3.85$, created by numerically integrating Eq. \eqref{CircPolBasisModel}. The solid lines of the scan show the natural evolution of the circulating field intensities, $|U_{1\rightarrow 4}|^2$, as the input power is gradually decreased. It can be seen that for the lower input power values of this scan the solid line (blue) mimics analogous results of the separate models described in the introduction; that is to say that for very low input intensities all four fields have equal intensities $|U_1|^2=|U_2|^2=|U_3|^2=|U_4|^2$, before the system displays a SSB bifurcation at a higher input intensity of around 0.82 and the four fields separate into two stable pairs of symmetric fields (cyan), with those pairings being either $|U_1|^2=|U_2|^2\;\&\;|U_3|^2=|U_4|^2$ or $|U_1|^2=|U_3|^2\;\&\;|U_2|^2=|U_4|^2$. This amounts to either a propagation direction or circular-polarization symmetry breaking respectively, with the type of SSB and the dominant and submissive roles of the two asymmetric pairs both being randomly assigned by the noise within the system. In this region, however, there are other possible symmetry broken solutions corresponding to only two fields having the symmetric intensity -- displayed by the dashed green lines in Fig.2. The remaining two fields are left asymmetric both to each other and to the symmetric pair, i.e. $|U_{i}|^2\neq |U_{j}|^2 = |U_{k}|^2 \neq |U_{l}|^2\;\&\;|U_{i}|^2\neq|U_{l}|^2$, where here the solo symmetric pair can be with $|U_1|^2=|U_4|^2$ or $|U_2|^2=|U_3|^2$. This indicates a symmetry retention of one of the diagonals of Table \ref{table:4Fields} with asymmetry between the remaining diagonal elements of the table. Contrary to the cyan line however, in this region the green solution line is unstable.
Although the cyan and green lines correspond to situations where the intensities of some of the field components remain equal to each other, at least two of the index exchanges that leave the system (4) invariant (symmetries) have now been broken. Unseen in the models of the separate systems, at even higher input intensities, at around 1.29, the system displays a second SSB bifurcation, leading to the situation of four completely asymmetric fields (yellow solid line), $|U_1|^2\neq |U_2|^2\neq |U_3|^2\neq |U_4|^2$. Finally, and somewhat surprisingly there is then a fourth region, caused by a ``partial'' symmetry-restoring bifurcation. In this region, the previously unstable solutions outlined above (dashed green line), describing a solo symmetric pair, becomes stable (solid green).

The dashed lines of Fig. \ref{fig:NestedSym} show stationary but unstable states of the system, still discoverable by numerical integration when we forcibly set, for example, $U_1=U_2=U_3=U_4$ (blue), $(U_1=U_2)\;\&\;(U_3=U_4)$ (cyan), and $U_1 = U_4$ (green).

In summary, we find four system states in relation to degrees of symmetry in the system: full symmetry, two symmetric pairs, one symmetric pair with the others asymmetric, and full asymmetry.

\begin{figure}
    \centering
    \includegraphics[width=8.5cm]{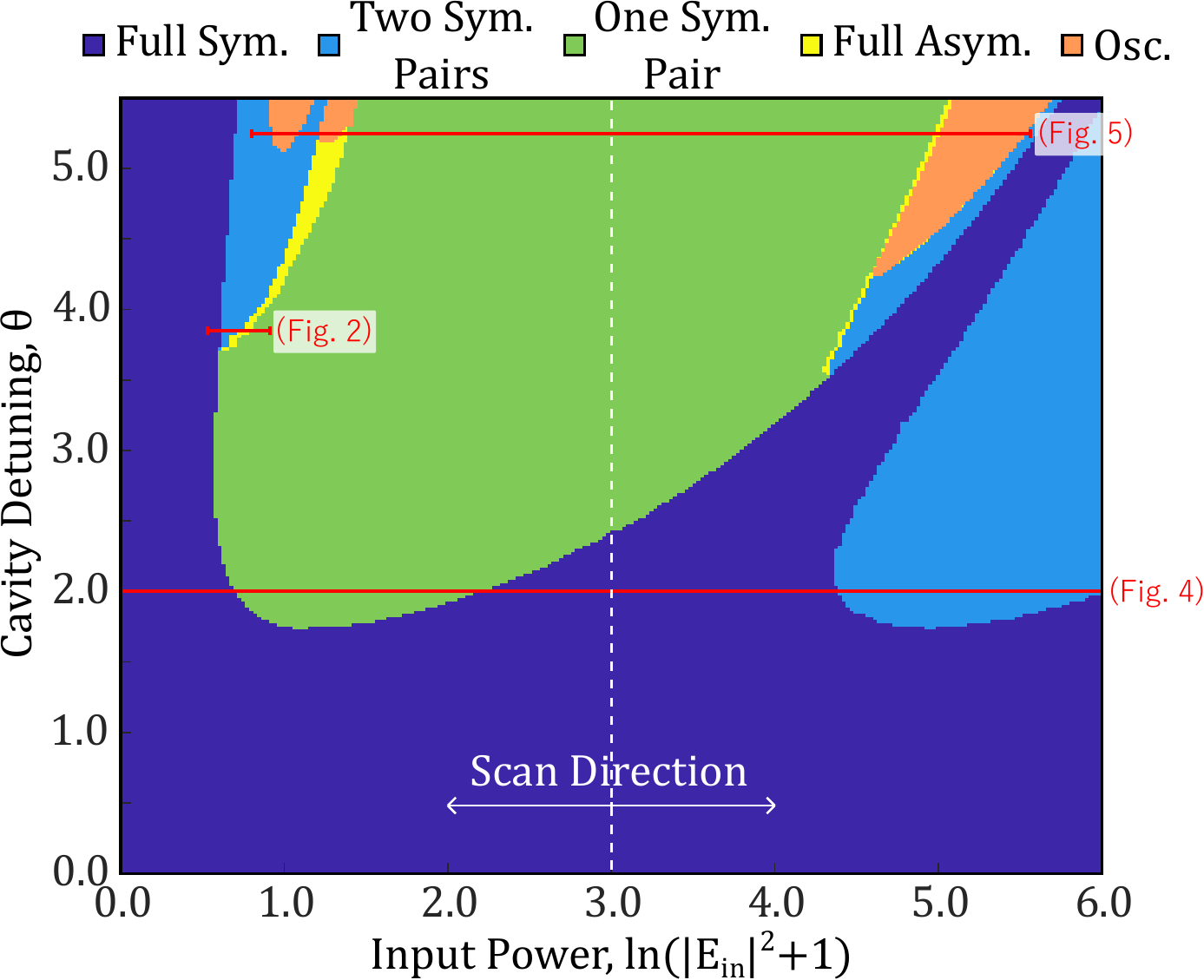}
    \caption{``Input power - cavity detuning'' parameter space. The dark blue region corresponds to fully symmetric stable solutions, the light blue region to two asymmetric pairs of symmetric fields, the green region to a single pair of symmetric fields, with the remaining two fields being both asymmetric to each other and to the symmetric pair, and finally the yellow region corresponds to four fully asymmetric fields. Also included is an orange overlay which indicates the regions where the fields are unstable to temporal oscillations. Red lines indicate the slices of the scan shown in Figs \ref{fig:NestedSym}, \ref{fig:DoubleSym} and \ref{fig:FullOScan} respectively.}
    \label{fig:RegionScan}
\end{figure}

\begin{figure}
    \centering
    \includegraphics[width=8.5cm]{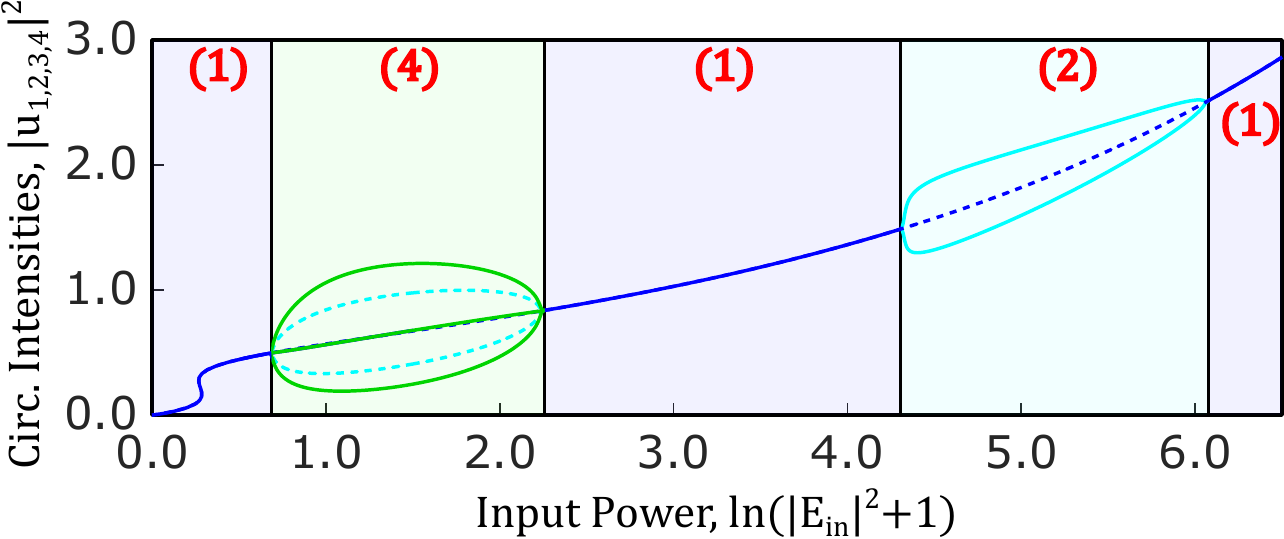}
    \caption{Input intensity scan of Eq. \eqref{CircPolBasisModel} for $\theta=2$. For this low detuning value we provide a much larger range of input intensity values than in Fig. \ref{fig:NestedSym}. This is to highlight a second symmetry broken region which emerges at high input intensity levels (light blue region on the RHS of Fig. \ref{fig:RegionScan}).}
    \label{fig:DoubleSym}
\end{figure}

To gain a better understanding of the possible states of Eq. \eqref{CircPolBasisModel}, we display in Fig. \ref{fig:RegionScan} the results of numerical scans in the parameter space of input power $\ln(|E_{in}|^2+1)$ and cavity detuning $\theta$, with the various field intensity relations described in the figure caption. It can be seen that the system's dynamics are very rich, and that specific relationships between field intensities, such as being fully asymmetric, are highly dependent on the input parameters. Regions of special interest in this figure are the light blue region, which emerges on the right-hand side of the plot, and the orange regions, indicating temporal oscillations, which emerge at high detuning values. First, to explain the onset and interest of the light blue regions, we provide an input intensity scan for $\theta=2$ in Fig. \ref{fig:DoubleSym}. Here we show, again with solid lines, the natural evolution of the circulating fields as we decrease the input power. Also shown with dashed lines, are the stationary but unstable solutions when we force $U_1=U_2=U_3=U_4$ (dashed blue) and $U_1=U_2, U_3=U_4$ (dashed cyan). At low input powers we see that with this four equation model, the states with a single symmetric pair are stable at the expense of the state with two asymmetric symmetric-pairs (analogous to the symmetry broken states of the  model of \cite{geddes1994polarisation}). Perhaps the most surprising result however is the second, entirely separate, symmetry broken regime which emerges at high input intensities which does not have an analogous region in the separate models despite existing solely between single SSB and symmetry-restoring bifurcations. Within this region lies the previously unstable asymmetric symmetric-pairs of fields. That is to say, in this high input intensity region the pairings $|U_1|^2=|U_4|^2\;\&\;|U_2|^2=|U_3|^2$ are stable, with the other possible combinations of pairings previously described now being unstable. This means that within our system any combination of two asymmetric symmetric-pairs are stable at some point in parameter space.

\section{\label{sec:Oscil}Oscillations, Chaotic Attractors, and Spontaneous Switching}

\begin{figure}
    \centering
    \includegraphics{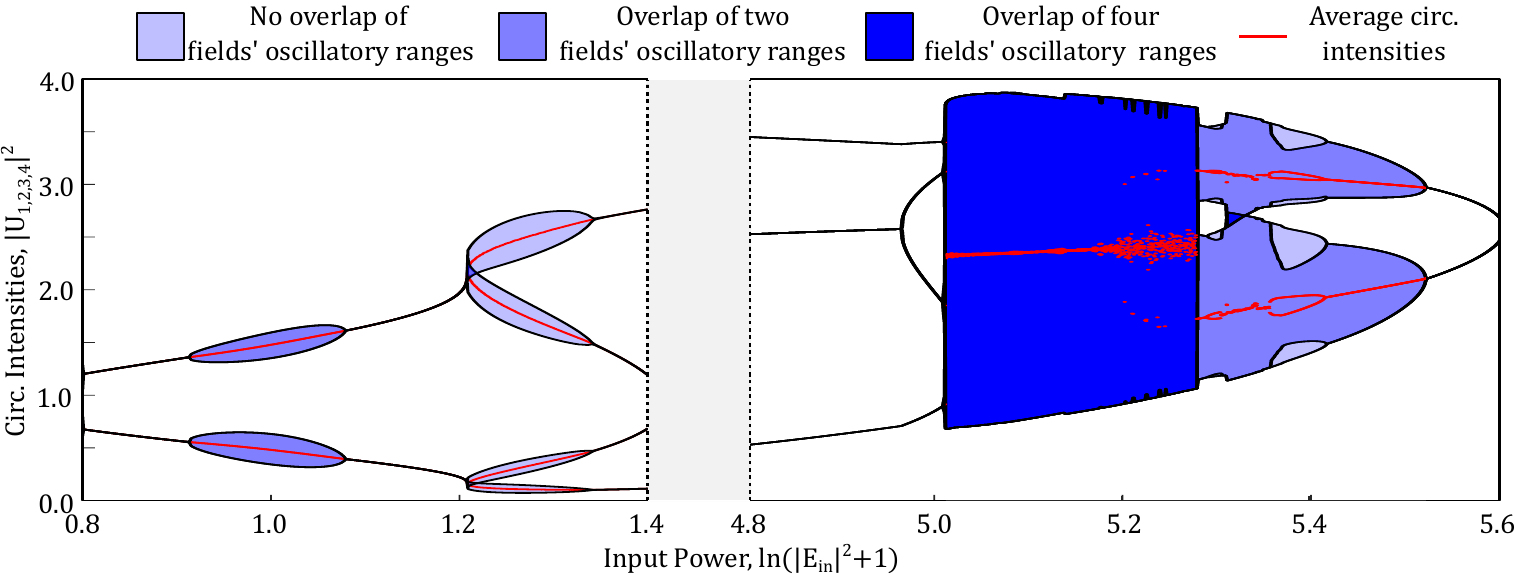}
    \caption{Field oscillation intensity ranges against input intensity for $\theta=5.25$. The solid black lines of the plot trace the maxima and minima of the four field's oscillatory ranges for the given input intensities (meaning up to eight black lines for some input intensities are possible). Regions shaded in different tones of blue indicate where oscillations are occurring. The tone of the blue shading itself indicates how many different fields overlap in their respective oscillation intensity ranges. The lightest blue indicates no overlap between any number of fields, the darkest shading implying all four field's oscillatory ranges overlap in this region, and finally the intermediate tone implies that only two of the four field's oscillatory ranges overlap in this region. The red lines track the averages of the four fields over many oscillations, four red lines are always present but may overlap where fields have globally symmetric intensity evolutions.}
    \label{fig:FullOScan}
\end{figure}

As hinted at by the orange overlay on Fig. \ref{fig:RegionScan}, the system described by Eq. \eqref{CircPolBasisModel} is susceptible to oscillations for some parameter regions. In Fig. \ref{fig:FullOScan} we display the full range of possible oscillatory behaviours for a selected value of the detuning $\theta=5.25$. In subsequent figures we display these behavioural regions in more detail.

\begin{figure}
    \centering
    \includegraphics[width=8.5cm]{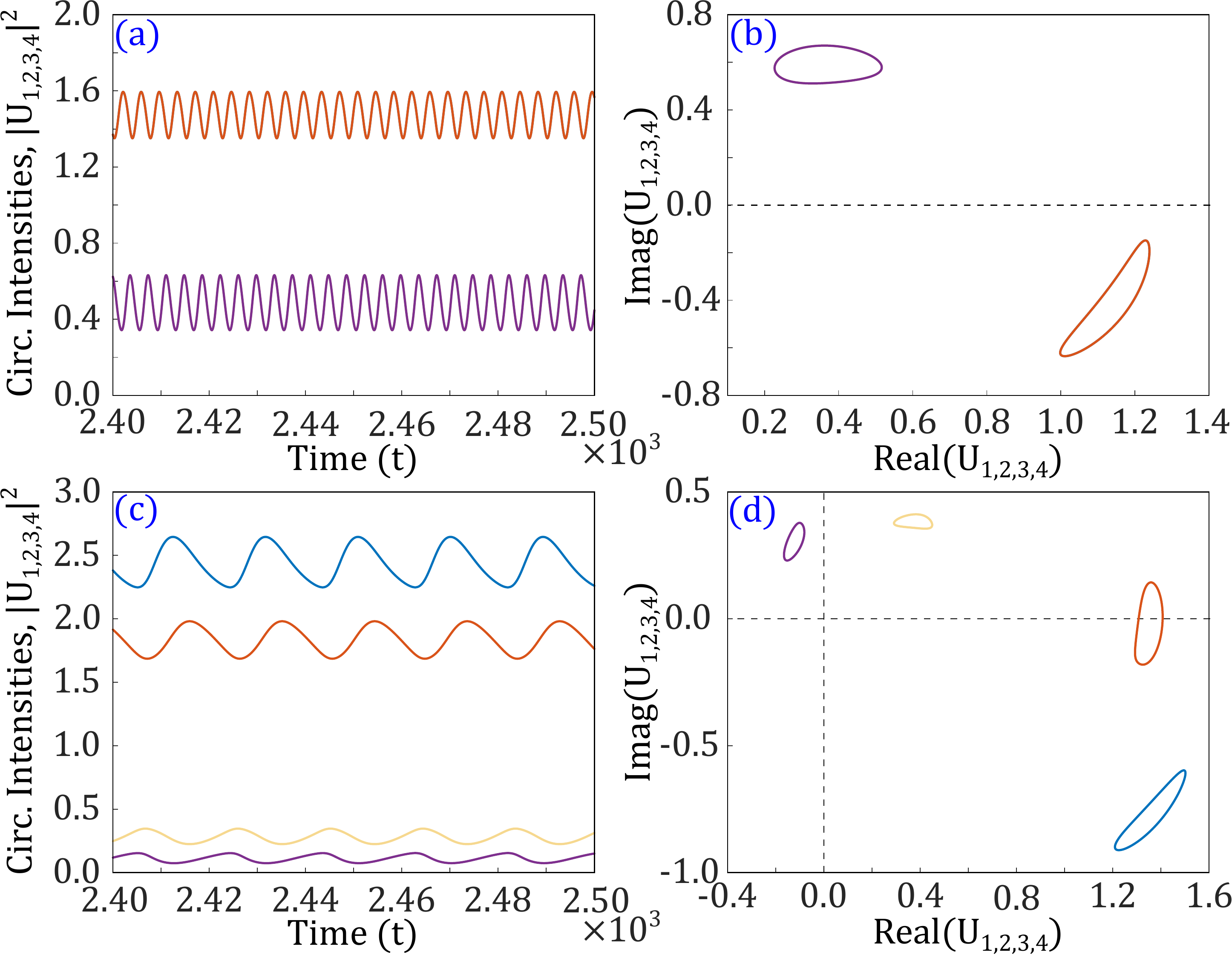}
    \caption{Field intensity evolutions over time (left panels), and their complex phase-space paths (right panels), for $\theta=5.25$, and for $|E_{in}|^2=1.0$ (top panels) and $|E_{in}|^2=1.25$ (bottom panels), computed by integrating Eq. \eqref{CircPolBasisModel}. Note that in panels (a,b) two pairs of fields evolve together causing two pairs of exact line overlaps, where as in panels (c,d) all four fields evolve asymmetrically to each other resulting in four distinct lines in each panel.}
    \label{fig:LowOsciEx}
\end{figure}

Focusing first on the LHS of Fig. \ref{fig:FullOScan}, in Fig. \ref{fig:LowOsciEx} we present the temporal evolutions of the intensities of the four fields and in the corresponding complex planes after transients have been discarded for $\theta=5.25$, and $|E_{in}|^2=1.0$ (top panels), $|E_{in}|^2=1.25$ (bottom panels). The examples provided in Fig. \ref{fig:LowOsciEx} are characteristic of the two orange regions on the far left of Fig. \ref{fig:RegionScan}, respectively, and show that these regions are quite different in nature. The first region displays oscillations where the four fields have split into two asymmetric symmetric-pairs, as described in the above section, (see Fig. \ref{fig:LowOsciEx}(a-b)), whereas in the second region all four fields oscillate in a fully asymmetric way with respect to one another (see Fig. \ref{fig:LowOsciEx}(c-d)).

\begin{figure}
    \centering
    \includegraphics[width=8.5cm]{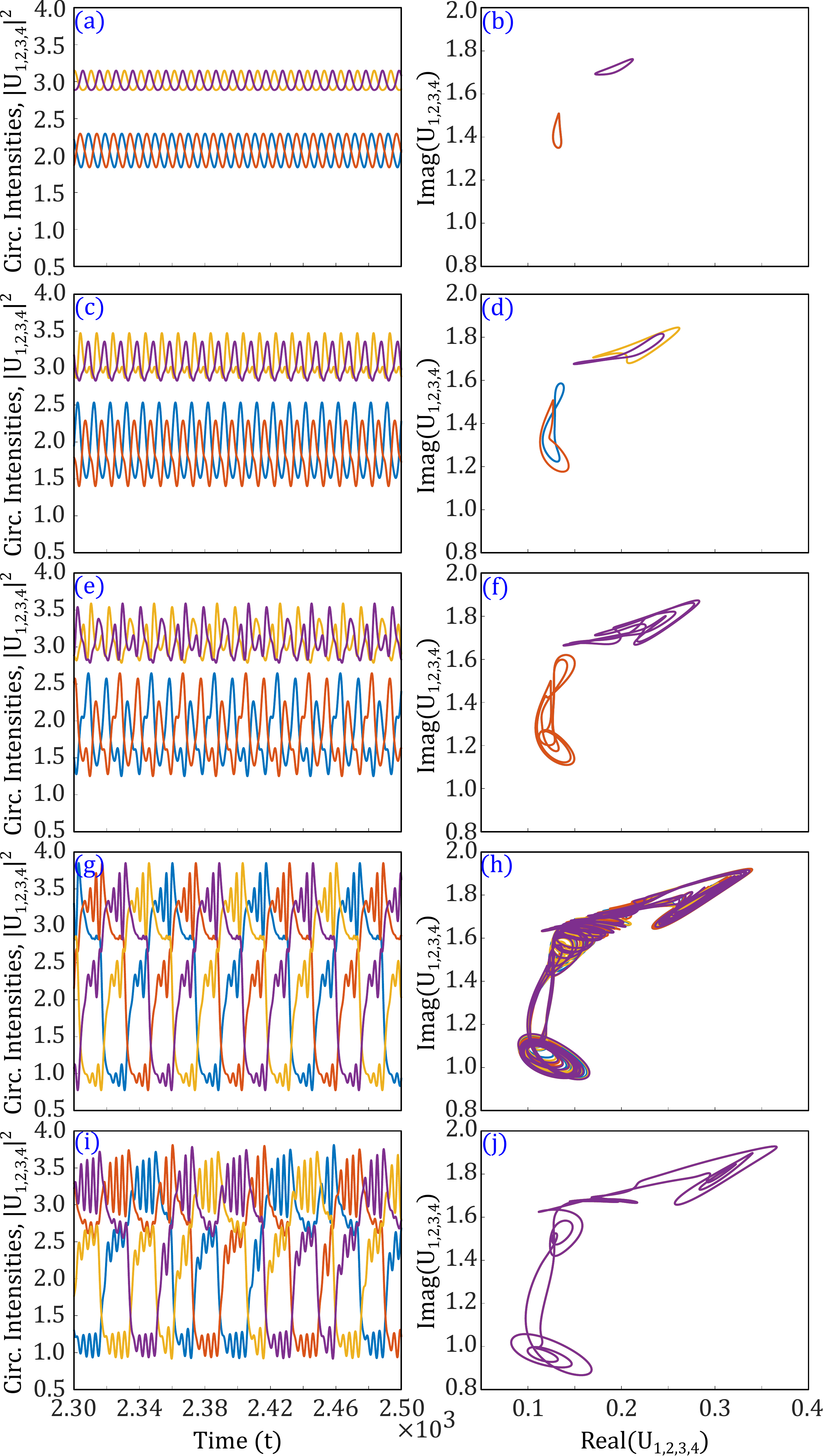}
    \caption{Field intensity evolutions over time (left panels), and their complex phase-space paths (right panels), for $\theta=5.25$, and for (a,b): $|E_{in}|^2=5.5$, (c,d): $|E_{in}|^2=5.4$, (e,f): $|E_{in}|^2=5.3525$, (g,h): $|E_{in}|^2=5.2$, and (i,j): $|E_{in}|^2=5.1$, computed by numerically numerically integrating Eq. \eqref{CircPolBasisModel}. For wider discussions please refer to main text.}
    \label{fig:OsciBehav2}
\end{figure}

Turning attention to the RHS of Fig. \ref{fig:FullOScan}, we now focus on the the range of field intensity oscillations for high input intensities. We show, in Fig. \ref{fig:OsciBehav2}, the distinct types of oscillations which are possible here.
The first type of oscillation occurs when two asymmetric symmetric-pairs of fields, again with the possible combinations of fields described in the above section, follow the same phase-space paths and have the same average intensities (see red lines of Fig. \ref{fig:FullOScan}). An example of this type of behaviour is displayed in Fig. \ref{fig:OsciBehav2} (a,b), for input intensity $E_{in}=5.5$. Due to the overlap of the phase space paths and the average intensities, this type of oscillation is very similar to those shown in Fig. \ref{fig:LowOsciEx} (a,b), which leads to this state retaining some of its underlying symmetry. We note, however, that unlike Fig. \ref{fig:LowOsciEx} (a,b) the fields here do not oscillate in phase and all four fields have fully asymmetric intensity evolutions. This type of behaviour provides an alternative way to obtain self-switching between dominant and suppressed fields in polarisation or counter-propagating systems, as discussed in Ref. \cite{woodley2021self}. In Ref. \cite{woodley2021self} the "periodic self-switching"-regions were extremely narrow, while here we observe very broad regions where periodic self-switching behaviour is possible.

In Fig. \ref{fig:FullOScan}, one observes a pale-blue region around $|E_{in}|^2=5.4$ beginning and ending with bifurcations in both the black (tracking the minima and maxima of the field intensity oscillations) and red (tracking the average of the field intensity oscillations) lines. These bifurcations beginning and ending the region are known as global symmetry breaking and global symmetry restoring bifurcations respectively, and correspond to a single symmetric attractor splitting into two attractors and to two attractors being the symmetric of each other and merging. In either case, the global bifurcations produce sudden and large changes to the morphology of the attractors and can be identified by the merging of attractor paths in the complex phase space. The types of oscillations which occur within these regions are characterised by the example oscillation traces shown in Fig. \ref{fig:OsciBehav2} (c,d), where $|E_{in}|^2=5.4$. Note that the previous pairs of fields which held residual average-intensity symmetry have split, and their complex-phase-space attractors have also split into two separate attractors - indicating a global symmetry breaking bifurcation. Ending these regions however, the average-intensity and complex-phase-space-path symmetries are restored, and the split attractors merge once again (a global symmetry restoring bifurcation),with an example of this shown in Fig. \ref{fig:OsciBehav2} (e,f) where $|E_{in}|^2=5.35$. 

There are however extended regions (seen where a single continuous red line is displayed in Fig. \ref{fig:FullOScan}) where all attractors merge perfectly, resulting in all four fields displaying symmetric average intensities and perfectly overlaid phase-paths Fig. \ref{fig:OsciBehav2} (i,j).

\section{Soliton-like Behaviour}

\begin{figure}
    \centering
    \includegraphics[width=8.5cm]{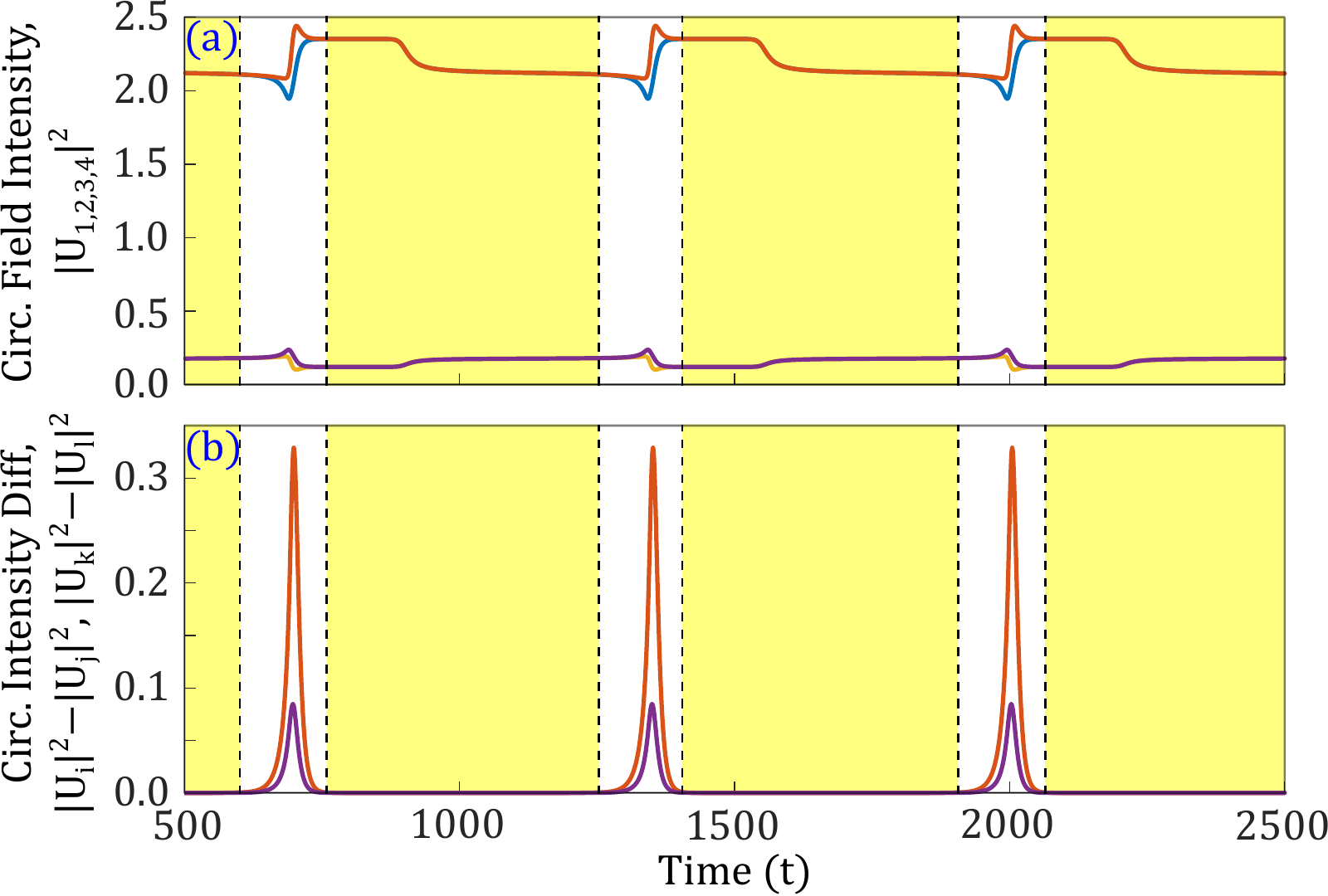}
    \caption{By taking the difference between the intensities of the two fields existing in close pairs (a) as they evolve over time, soliton like structures (b) can be observed with long delays between their production. The example displayed is for an input laser intensity of $|E_{in}|^2\approx 1.21$ and detuning $\theta=5.25$.}
    \label{fig:DelayedSol}
\end{figure}

In addition to oscillatory regimes, we found solutions that exhibit soliton-like behavior on the slow time scale in the region where all four fields oscillate with separate values of their intensities as one approaches the region's low-input-intensity limit, at approximately $|E_{in}|^2=1.21$ in Fig. \ref{fig:FullOScan}. As shown in Fig. \ref{fig:DelayedSol}, there is the possibility to generate soliton-like structures on a time scale of many round trips of the resonator with long delays between each structure generation. The soliton-like structures can be observed by taking the difference between two of the circulating fields, as shown in Fig. \ref{fig:DelayedSol}(b), with possible pairings of the top two and bottom two fields being $|U_1|^2=|U_2|^2\;\&\;|U_3|^2=|U_4|^2$ or $|U_1|^2=|U_3|^2\;\&\;|U_2|^2=|U_4|^2$. i.e the system is capable of periodically almost gaining and then rapidly losing either of the polarization or propagation-direction symmetries in this region.

\begin{figure}
    \centering
    \includegraphics[width=8.5cm]{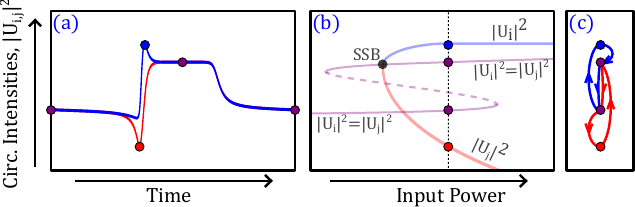}
    \caption{(a): Intensity evolutions which can cause soliton-like structures when subtracted. (b): Theorised HSS diagram explaining the field dynamics producing (a). (c) Visual representation of how the fields oscillate between the various HSS that we theorise are possible in (b).}
    \label{fig:Osc.Exp1}
\end{figure}

Focusing on only, for example, the top two fields of Fig. \ref{fig:DelayedSol}, in Fig. \ref{fig:Osc.Exp1}(b) we show our theorised HSS. It contains an optical bistability cycle of the symmetric state, say $|U_1|^2=|U_2|^2$, coexisting with asymmetric solutions, all of which are unstable to oscillations. The observed intensity evolutions of Fig. \ref{fig:DelayedSol}(a) can then be explained by the following consideration: focusing on one period of the intensity oscillation in Fig. \ref{fig:Osc.Exp1}(a), the two fields begin with almost perfect polarization or propagation-direction symmetry on the lower branch of the optical bistability of Fig. \ref{fig:Osc.Exp1}(b), purple dot. This HSS is metastable so that the system initially evolves away from this symmetric point and is attracted to the two asymmetric solutions, blue and red dots respectively of panel (b), leading to an increasingly large asymmetry between, as an example, the intensities of the two polarization components of the field. Since we predict the asymmetric HSS to be unstable to oscillations however, the system does not settle on the asymmetric values, and instead it is drawn to a predicted attractive but metastable symmetric solution which lies on the upper branch of the symmetric optical bistability. Unable to find a stable solution even here however, the system continues to oscillate where it is attracted to the original symmetric, but metastable, solution on the lower branch of the HSS, and the cycle then repeats. This process is summarised in panel (c) with arrowed paths.

The proposed explanation in terms of metastable states on an optical bistable cycle and unstable asymmetric branches is further supported by similar observations in the separate systems \cite{kaplan1982directionally}, supplemented with the, required, nested spontaneous symmetry breaking bifurcations -- the primary result of this paper.

\section{\label{sec:Conc}Conclusions}
We presented a model for Kerr ring resonators with two counter-propagating input beams with two orthogonally polarized components for each beam. The model predicts that the physical system will be able to display a vast number of novel behaviours while simultaneously also being able to mimic simpler systems if required.

We showed that the system is capable of multiple, nested and isolated, symmetry breaking bifurcations and restoring events; capable of displaying full symmetry, partial-symmetry breaking and restoring, and even total asymmetry, with full freedom of choice for each of the field roles. This allows for, for example, the realisation of polarisation symmetry breaking in a single direction while polarization symmetry is maintained in the other, or for directional-symmetry breaking for fields with a certain polarization, while directional symmetry is maintained for the other polarization, useful for polarization-dependent isolators and circulators.

Turning attention to the stability of the system and its susceptibility to oscillations, we completed a behavioural scan of the system for a range of parameters, which revealed a number of regions where the system was unstable and susceptible to a range of oscillatory behaviours -- again with various degrees of symmetry broken or retained. We showed that multiple chaotic attractors can merge as input parameters are varied, and that this leads to not only two-field self-switching but also the complex dynamical behaviour of four-field self-switching, brought about by a global symmetry restoring bifurcation. Contrary to to the self-switching dynamics of the separate systems, which occurred for only small windows of input parameters, our self-switching dynamics were shown to be maintained for large input ranges.

Atop these novel field behaviours, the system further provides the option of producing, on a single device, the field behaviours of the two separate systems, counter-propagation and orthogonal polarizations. This is to say, it can produce the behaviours of two counter propagating fields -- lending itself to applications such as enhancing the Sangac effect for use in rotation sensors \cite{kaplan1981enhancement}, gyroscopes \cite{silver2017nonlinear}, and elsewhere, and in the realisation of all-optical components, such as isolators and circulators \cite{del2018microresonator}, while also being able to produce the behaviours of the system with two orthogonal polarisation components, allowing for applications such as acting as a Kerr polarisation controller \cite{moroney2022kerr}, potentially supporting temporal cavity solitons for the generation of frequency combs \cite{xu2021spontaneous, xu2022breathing}, or even for use in optical neural networks used in artificial intelligence applications or for systems for quantum information processing. The benefit of this combined system however should be obvious -- it can achieve the behaviours of both systems simultaneously, all while taking place within one resonator. For industrial applications this has the benefit of saving both space and potentially money, and should be easily achievable with current resonator or fibre loop technology. There would also be applications of the physical system in high speed telecommunication systems, particularly for polarization mode multiplexing and for the possibility of having counter-propagating light in telecoms systems. For these reasons, physical realisation of these devices can benefit from the wide range of predicted behaviours displayed here.

\begin{backmatter}
\bmsection{Funding}
LH acknowledges funding provided by the CNQO group within the Department of Physics at the University of Strathclyde, and the "Saltire Emerging Researcher" scheme through the Scottish University's Physics Alliance (SUPA) and provided by the Scottish Government and Scottish Funding Council. This work was further supported by the European Union’s H2020 ERC Starting Grants 756966, the Marie Curie Innovative Training Network “Microcombs” 812818 and the Max Planck Society.
\bmsection{Disclosures}
The authors declare no conflicts of interest.
\bmsection{Data Availability Statement}
\end{backmatter}

\bibliography{Ref}

\end{document}